\documentclass[article]{jss}\usepackage[]{graphicx}\usepackage[]{color}
\makeatletter
\def\maxwidth{ %
  \ifdim\Gin@nat@width>\linewidth
    \linewidth
  \else
    \Gin@nat@width
  \fi
}
\makeatother

\definecolor{fgcolor}{rgb}{0.345, 0.345, 0.345}
\newcommand{\hlnum}[1]{\textcolor[rgb]{0.686,0.059,0.569}{#1}}%
\newcommand{\hlstr}[1]{\textcolor[rgb]{0.192,0.494,0.8}{#1}}%
\newcommand{\hlcom}[1]{\textcolor[rgb]{0.678,0.584,0.686}{\textit{#1}}}%
\newcommand{\hlopt}[1]{\textcolor[rgb]{0,0,0}{#1}}%
\newcommand{\hlstd}[1]{\textcolor[rgb]{0.345,0.345,0.345}{#1}}%
\newcommand{\hlkwb}[1]{\textcolor[rgb]{0.69,0.353,0.396}{#1}}%
\newcommand{\hlkwc}[1]{\textcolor[rgb]{0.333,0.667,0.333}{#1}}%
\newcommand{\hlkwd}[1]{\textcolor[rgb]{0.737,0.353,0.396}{\textbf{#1}}}%

\usepackage{framed}
\makeatletter
\newenvironment{kframe}{%
 \def\at@end@of@kframe{}%
 \ifinner\ifhmode%
  \def\at@end@of@kframe{\end{minipage}}%
  \begin{minipage}{\columnwidth}%
 \fi\fi%
 \def\FrameCommand##1{\hskip\@totalleftmargin \hskip-\fboxsep
 \colorbox{shadecolor}{##1}\hskip-\fboxsep
     \hskip-\linewidth \hskip-\@totalleftmargin \hskip\columnwidth}%
 \MakeFramed {\advance\hsize-\width
   \@totalleftmargin\z@ \linewidth\hsize
   \@setminipage}}%
 {\par\unskip\endMakeFramed%
 \at@end@of@kframe}
\makeatother

\definecolor{shadecolor}{rgb}{.97, .97, .97}
\definecolor{messagecolor}{rgb}{0, 0, 0}
\definecolor{warningcolor}{rgb}{1, 0, 1}
\definecolor{errorcolor}{rgb}{1, 0, 0}
\newenvironment{knitrout}{}{} 

\usepackage{alltt}

\usepackage{thumbpdf,lmodern}

\usepackage{framed}

\usepackage{amsmath}
\usepackage{subfig}
\usepackage{float}

\author{Anita K Nandi\\University of Oxford
   \And Tim CD Lucas\\University of Oxford
   \And Rohan Arambepola\\University of Oxford
   \AND Peter Gething\\Telethon Kids Institute \\ Curtin University
   \And Daniel J Weiss\\University of Oxford}
\Plainauthor{Anita K Nandi, Tim CD Lucas, Rohan Arambepola, Peter Gething, Daniel J Weiss}

\title{disaggregation: An \proglang{R} Package for Bayesian Spatial Disaggregation Modelling}
\Plaintitle{disaggregation: An R Package for Bayesian Spatial Disaggregation Modelling}
\Shorttitle{Disaggregation Modelling in \proglang{R}}

\Abstract{
  Disaggregation modelling, or downscaling, has become an important discipline in epidemiology. Surveillance data, aggregated over large regions, is becoming more common, leading to an increasing demand for modelling frameworks that can deal with this data to understand spatial patterns. Disaggregation regression models use response data aggregated over large heterogenous regions to make predictions at fine-scale over the region by using fine-scale covariates to inform the heterogeneity. This paper presents the \proglang{R} package \emph{disaggregation}, which provides functionality to streamline the process of running a disaggregation model for fine-scale predictions.
}

\Plainkeywords{Bayesian spatial modelling, disaggregation modelling, TMB}

\Address{
  Anita Nandi\\
  Malaria Atlas Project\\
  Big Data Institute\\
  Old Road Campus\\
  University of Oxford\\
  Oxford, UK\\
  E-mail: \email{anita.nandi@bdi.ox.ac.uk}
}
\IfFileExists{upquote.sty}{\usepackage{upquote}}{}
\begin{document}



\section{Introduction} \label{sec:intro}

Methods for estimating high-resolution risk maps from aggregated response data over large spatial regions are 
becoming increasingly sought after in disease risk mapping~\citep{li2012log, diggle2013spatial, wilson2017pointless}, especially in malaria mapping~\citep{sturrock2014fine, sturrock2016mapping}. 
Disaggregation regression, first applied in species distribution modelling in ecology~\citep{keil2013downscaling, barwell2014can} has now become an important method in disease risk mapping \citep{falciparum2019, vivax2019}. 
The aggregation of response data over large heterogenous regions is problematic for making fine-scale predictions, as relationships learned between variables at one spatial scale may not hold at other scales \citep{wakefield2006health}. However, by using fine-scale information from related covariates we can inform the hetrogeneity of the response variable of interest within the aggregated area.

Disaggregation modelling is unorthodox as the predictions are at a different scale to the response data, i.e. the number of rows in the covariate data is different to that of the response data. The spatial modelling software package, \pkg{INLA}~\citep{rue2009approximate}, or integrated nested Laplace approximation, has been shown to very useful in a wide variety of circumstances, however it is not flexible enough for the unorthodox nature of the disaggregation problem except in the special case of the linear link function \citep{wilson2017pointless}. Disaggregation models can be implemented in \pkg{TMB}~\citep{tmb2016}, or template model builder, with a lot of flexibility, however the data manipulation required to format the model objects and construct the model definition in C++ is non-trivial. The \pkg{disaggregation} package allows this process to be streamlined, to make it easy for the user to run disaggregation models at the expense of some flexibility.



\section{Disaggregation modelling} \label{sec:model}

Suppose we have response data, $y_i$, for $N$ polygons, which corresponds to count data for the property of interest within that polygon. 
The process that is being counted occurs in continuous space that we model as a high-resolution, square lattice for convenience.
The data, $y_i$,  are assumed to be created by the aggregation of the counts over the polygon, i.e. the count data of the polygon is given by the sum of the count data for all the pixels within that polygon. 
The rate is defined such that the number of cases in this pixel can be calculated by multiplying the rate by the aggregation raster. For example, in epidemiology, we may have as our response the number of people that contract a certain disease in a given period of time (case incidence). Our rate would be the incidence rate of cases per population, where the aggregation raster corresponds to population. 

For the disaggregation model we model the rate at pixel level, with the likelihood for the observed data given by aggregating these pixel level rates. The rate in pixel $j$ of polygon $i$ at location $s_{ij}$ is given by:
\begin{equation} \label{eq:lp}
\textrm{link}(\textrm{rate}_{ij}) = \beta_0 + \beta X_{ij} + \textrm{GP}(s_{ij}) + u_i 
\end{equation}
where $\beta$ are regression coefficients, $X_{ij}$ are covariate values, GP is a Gaussian random field and $u_i$ is a polygon-specific iid effect. The user-defined link function is typically log, identity and logit for Poisson, Normal and Binomial likelihoods, respectively.  The Gaussian random field has a Matern covariance function parameterised by $\rho$, the range (approximately the distance beyond which correlation is less than 0.1), and $\sigma$, the marginal standard deviation. As we are working in a Bayesian setting, each of the model parameters and hyperparameters are given a prior, which is discussed later. 

The pixel predictions are then aggregated to the polygon level using the weighted sum (via the aggregation raster $a_{ij}$):

\begin{equation} \label{eq:agreggate1}
\textrm{cases}_i = \sum_{j=1}^{N_i} a_{ij}\textrm{rate}_{ij}
\end{equation}
\begin{equation} \label{eq:agreggate2}
\textrm{rate}_i = \frac{\textrm{cases}_{i}}{\sum_{j=1}^{N_i} a_{ij}}
\end{equation}
The different likelihoods correspond to slightly different models ($y_i$ is the response count data):

\begin{itemize}
\item \textbf{Poisson}
\begin{equation} \label{eq:poisson}
y_i \sim \textrm{Pois}(\textrm{cases}_i)
\end{equation}

\item {\textbf{Gaussian}
\begin{equation} \label{eq:gaussian}
y_i \sim \textrm{Normal}(\textrm{cases}_i, \sigma_i)
\end{equation}
Here $\sigma_i = \sigma \sqrt{\sum_j a_{ij}^2}$, where  $\sigma$ is the pixel-level dispersion (a parameter learnt by the model) }

\item \textbf{Binomial}
\begin{equation} \label{eq:binomial}
y_i \sim \textrm{Binomial}(M_i, \textrm{rate}_i)
\end{equation}

\end{itemize}
In the example of disease mapping, Poisson or Gaussian likelihoods could be used when the quantity observed, $y_i$, is the total number of cases in a given polygon. The Binomial model could be used when $y_i$ is the prevalence of a disease in a sample of $M_i$ people in the polygon.

\subsection{Priors}

For each of the model parameters and hyperparameters we specify priors. The regression parameters and intercept are given Gaussian priors, where the default priors are $\beta_0 \sim N(0, 2)$ and $\beta_i \sim N(0, 0.4)$. A penalised complexity prior is placed on the scale and range parameters of the random field \citep{fuglstad2018constructing} such that
\begin{align}
    P(\rho < \rho_{min}) &= \rho_{prob} \\
    P(\sigma > \sigma_{max}) &= \sigma_{prob}
\end{align}
where the values $\rho_{min}, \,\rho_{prob}, \,\sigma_{max}, \,\sigma_{prob}$ are set by the user. This prior shrinks the field towards a base model with zero variance and infinite range, in other words regularising towards a flatter field with smaller magnitude. The polygon-specific effects $u_1,...,u_N$ have Gaussian priors centered at 0 with standard deviation $\sigma_{u}$ (where the precision $\tau_{u} = 1/\sigma_{u}^2$). A penalised complexity prior \citep{simpson2017penalising} is placed on $\sigma_{u}$ such that 
\begin{equation}
P(\sigma_{u} > \sigma_{u, max}) = \sigma_{u, prob}
\end{equation}
which shrinks towards a base model of no polygon-specific effect. The $\sigma_{u, max}$ and $\sigma_{u, prob}$ are set by the user.

For models that use a Gaussian likelihood, a log gamma prior is set on the log of the precision, $\log\tau \sim \log\Gamma(\text{shape} = 1, \text{rate} = 5 \times 10^{-5})$, to regularise the dispersion ($\sigma = 1/\sqrt\tau$) to take low values. This is the same as the prior set by INLA for the dispersion of the normal likelihood.

\section{TMB implementation} \label{sec:implementation}

The \pkg{disaggregation} package is built on template model builder (\pkg{TMB})~\cite{tmb2016}, which is a tool for flexibly building complex models based on C++. \pkg{TMB} combines the packages CppAD~\citep{bell2012cppad}, for automatic differentiation in C++, Eigen~\citep{guennebaud2010eigen}, a C++ library for linear algebra, and CHOLMOD~\citep{chen2008algorithm}, a package for efficient computation of sparse matrices. The use of these packages allows an efficient implementation of the automatic Laplace approximation~\citep{skaug2006automatic} with exact derivatives which gives an approximation to the Bayesian posterior. \pkg{TMB} calculates first and second order derivatives of the objective function using automatic differentiation~\citep{griewank2008evaluating}.

The \pkg{disaggregation} package contains a C++ function that defines the model and computes the joint likelihood as a function of the parameters and the random effects, in the format expected by \pkg{TMB}. The \pkg{TMB} package then calculates estimates of both parameters and random effects using the Laplace approximation for the likelihood. Evaluation of the objective function and its derivatives is performed via \proglang{R}.

\section{Package usage} \label{sec:code}

In this section we show how to use the \pkg{disaggregation} package using a dataset of aggregated malaria case counts across Madagascar. Malaria is an infectious disease caused by parasites of the \emph{Plasmodium} group, transmitted by \emph{Anopheles} mosquitoes. Malaria transmission is therefore closely related to mosquito and parasite development. Environmental factors such as temperature, rainfall and elevation have been shown to have signifcant effect on mosquito survival and development; in general mosquitoes favour warm, humid environments with moderate rainfall. Therefore, such environmental covariates would be useful in a malaria disaggregation model to inform fine-scale distributions. In this model we use the environmental covariates of mean land surface temperature (LST), variation in LST, elevation, and enhanced vegetation index (EVI) to inform spatial heterogeneity in malaria risk. These covariates are obtained from the Moderate Resolution Imaging Spectroradiometer (MODIS), which provides many measurements over the entire Earth's surface (\url{https://modis.gsfc.nasa.gov/data/}). 
Malaria incidence rate is often given per thousand cases and given the term annual parasite index (API).

The latest version of \pkg{disaggregation} should always be available from the Comprehensive R Archive Network (CRAN) at \href{http://CRAN.R-project.org/package=disaggregation}{http://CRAN.R-project.org/package=disaggregation}. Run the following commands to install and load the package.

\begin{knitrout}
\definecolor{shadecolor}{rgb}{0.969, 0.969, 0.969}\color{fgcolor}\begin{kframe}
\begin{alltt}
\hlstd{devtools}\hlopt{::}\hlkwd{install.packages}\hlstd{(}\hlstr{"disaggregation"}\hlstd{)}
\hlkwd{library}\hlstd{(disaggregation)}
\end{alltt}
\end{kframe}
\end{knitrout}

We then read in the data to use in the \pkg{disaggregation} package.

\begin{knitrout}
\definecolor{shadecolor}{rgb}{0.969, 0.969, 0.969}\color{fgcolor}\begin{kframe}
\begin{alltt}
\hlkwd{library}\hlstd{(raster)}

\hlstd{shapes} \hlkwb{<-} \hlkwd{shapefile}\hlstd{(}\hlstr{'data/shapes/mdg_shapes.shp'}\hlstd{)}
\hlstd{population_raster} \hlkwb{<-} \hlkwd{raster}\hlstd{(}\hlstr{'data/population.tif'}\hlstd{)}
\hlstd{covariate_stack} \hlkwb{<-} \hlkwd{getCovariateRasters}\hlstd{(}\hlstr{'data/covariates'}\hlstd{,}
                                       \hlkwc{shape} \hlstd{= population_raster)}
\end{alltt}
\end{kframe}
\end{knitrout}

The main functions are \code{prepare_data}, \code{fit_model} and \code{predict}.

Firstly, we use the \code{prepare_data} function to setup all the data in the format needed in the disaggregation modelling. This function performs various data manipulation tasks to create objects that are necessary for fitting the model. The required input data for the \code{prepare_data} function are a SpatialPolygonsDataFrame containing the response data and a RasterStack of covariates to be used in the model. The variable names in the SpatialPolygonsDataFrame of the response count data and the polygon ID are defined by the user.

An optional aggregation raster can be provided. The aggregation raster defines how the pixels within each polygon are aggregated. The disaggregation model performs a weighted sum of the pixel prediction, weighted by the pixel values in the aggregation raster. In this case our pixel predictions are malaria incidence rate, so we use the population raster to aggregate pixel incidence rate by summing the number of cases (rate weighted by population). If no aggregation raster is provided a uniform distribution is assumed, i.e. the pixel predictions are aggregated to polygon level by summing the pixel values unaltered.

The values of the covariates (as well as the aggregation raster, if given) are extracted at each pixel within the polygons and stored as a data.frame with a row for each pixel and a column for each covariate (\code{parallelExtract} function). The extraction of each covariate is performed in parallel over the number of cores defined by the argument \code{ncores}. The values extracted from the aggregation raster are returned as an array of values, one for each pixel.

In order to know which pixels (i.e. which rows) are contained in each polygon, a matrix is constructed that contains the start and end pixel index for each polygon (\code{getStartendindex} function). Additionally, an INLA mesh is built to use for the spatial field (\code{build_mesh} function). The \code{mesh.args} argument allows the user to supply a list of INLA mesh parameters to control the mesh used for the spatial field. The mesh can take several minutes to construct, so to prepare the data without buidling the mesh the user can set the \code{makeMesh} flag to FALSE. However, it would then not be possible to fit the disaggregation model without the mesh.

If there are any NAs in the response or covariate data within the polygons the \code{prepare_data} method will return an error. This can be dealt with using the \code{na.action} flag, which is automatically off.  Ideally the NAs in the data would be dealt with by the user beforehand, however, setting na.action = TRUE will automatically deal with NAs. It removes any polygons that have NAs as a response, sets any aggregation pixels with NA to zero and sets covariate NAs pixels to the median value for that covariate across all polygons.

\begin{knitrout}
\definecolor{shadecolor}{rgb}{0.969, 0.969, 0.969}\color{fgcolor}\begin{kframe}
\begin{alltt}
\hlstd{dis_data} \hlkwb{<-} \hlkwd{prepare_data}\hlstd{(}\hlkwc{polygon_shapefile} \hlstd{= shapes,}
                         \hlkwc{covariate_rasters} \hlstd{= covariate_stack,}
                         \hlkwc{aggregation_raster} \hlstd{= population_raster,}
                         \hlkwc{mesh.args} \hlstd{=} \hlkwd{list}\hlstd{(}\hlkwc{max.edge} \hlstd{=} \hlkwd{c}\hlstd{(}\hlnum{0.7}\hlstd{,} \hlnum{8}\hlstd{),}
                                          \hlkwc{cut} \hlstd{=} \hlnum{0.05}\hlstd{,}
                                          \hlkwc{offset} \hlstd{=} \hlkwd{c}\hlstd{(}\hlnum{1}\hlstd{,} \hlnum{2}\hlstd{)),}
                         \hlkwc{id_var} \hlstd{=} \hlstr{'ID_2'}\hlstd{,}
                         \hlkwc{response_var} \hlstd{=} \hlstr{'inc'}\hlstd{,}
                         \hlkwc{na.action} \hlstd{=} \hlnum{TRUE}\hlstd{,}
                         \hlkwc{ncores} \hlstd{=} \hlnum{8}\hlstd{)}
\end{alltt}
\end{kframe}
\end{knitrout}

We can see a summary of the data, using the generic \code{summary} function, and \code{plot} the data. The summary function returns information on how many pixels and polygons the data contains, how many pixels in the smallest and largest polygons and a summary of the covariate data. The \code{plot} functions plots a map of the polygon response data, the covariate rasters and the INLA mesh, as shown in Figure~\ref{fig:subplotdata}.

\begin{knitrout}
\definecolor{shadecolor}{rgb}{0.969, 0.969, 0.969}\color{fgcolor}\begin{kframe}
\begin{alltt}
\hlkwd{summary}\hlstd{(dis_data)}
\end{alltt}
\begin{verbatim}
#> They data contains 109 polygons and 28892 pixels
#> The largest polygon contains 867 pixels and the smallest polygon contains 1 pixels
#> There are 4 covariates
#> 
#> Covariate summary:
#>    Elevation              EVI              LSTmean        
#>  Min.   :-1.174670   Min.   :-2.70028   Min.   :-3.33993  
#>  1st Qu.:-0.848255   1st Qu.:-0.73278   1st Qu.:-0.73083  
#>  Median :-0.257381   Median :-0.36756   Median : 0.23743  
#>  Mean   : 0.009915   Mean   : 0.01077   Mean   : 0.01572  
#>  3rd Qu.: 0.713218   3rd Qu.: 0.59588   3rd Qu.: 0.83374  
#>  Max.   : 4.585433   Max.   : 3.14432   Max.   : 2.09070  
#>      LSTsd         
#>  Min.   :-2.92708  
#>  1st Qu.:-0.65033  
#>  Median :-0.09492  
#>  Mean   : 0.03608  
#>  3rd Qu.: 0.67376  
#>  Max.   : 3.58093
\end{verbatim}
\end{kframe}
\end{knitrout}
\begin{knitrout}
\definecolor{shadecolor}{rgb}{0.969, 0.969, 0.969}\color{fgcolor}\begin{kframe}
\begin{alltt}
\hlkwd{plot}\hlstd{(dis_data)}
\end{alltt}
\end{kframe}
\end{knitrout}
\begin{knitrout}
\definecolor{shadecolor}{rgb}{0.969, 0.969, 0.969}\color{fgcolor}\begin{figure}[H]
\subfloat[\label{fig:subplotdata1}]{\includegraphics[width=0.5\linewidth]{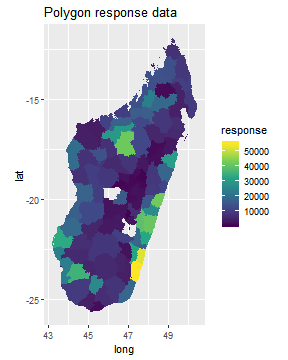} }
\subfloat[\label{fig:subplotdata2}]{\includegraphics[width=0.5\linewidth]{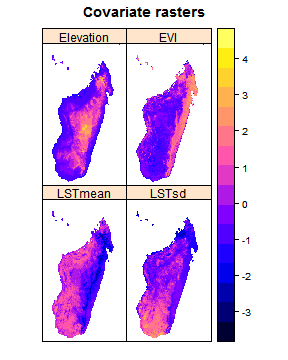} }\newline
\subfloat[\label{fig:subplotdata3}]{\includegraphics[width=0.5\linewidth]{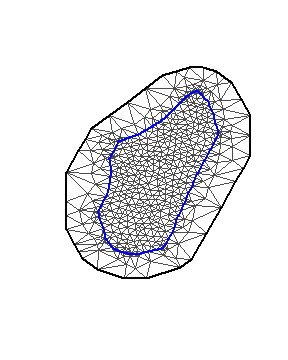} }\caption[Maps of Madagascar showing the data used in the disaggregation model]{Maps of Madagascar showing the data used in the disaggregation model. These plots are produced when calling the plot function on the disag.data object. The plots show: (a) number of malaria cases in each administrative unit, (b) maps of the four covariates used in the model; and (c) inla.mesh object that will be used to make the spatial field.}\label{fig:subplotdata}
\end{figure}

\end{knitrout}

The \code{prepare_data} function returns an object of class \code{disag.data}, which is designed to be used directly in the \code{fit_model} function.

Now we can fit the disaggregation model using \code{fit_model}. Here we use a Poisson likelihood for the incident count data with a log link function. Options for the likelihood are Gaussian, Poisson and binomial, and options for the link function are logit, log and identity. The spatial field and iid effect are components of the model by default, they can be turned off using the \code{field} and \code{iid} flags. We specify all of the priors for the model in a single list. Hyperpriors for the field are given as penalised complexity priors - specify $\rho_{min}$ and $\rho_{prob}$ for the range of the field, where $P(\rho < \rho_{min}) = \rho_{prob}$, and $\sigma_{min}$ and $\sigma_{prob}$ for the variation of the field, where $P(\sigma > \sigma_{min}) = \sigma_{prob}$. Similarly, the user specifies penalised complexity priors for the iid effect.
The \code{iterations} argument specifies the maximum number of iterations the model can run for to find an optimal point. In order to print more verbose output the user can set the \code{silent} argument to FALSE.

\begin{knitrout}
\definecolor{shadecolor}{rgb}{0.969, 0.969, 0.969}\color{fgcolor}\begin{kframe}
\begin{alltt}
\hlstd{fitted_model} \hlkwb{<-} \hlkwd{fit_model}\hlstd{(}\hlkwc{data} \hlstd{= dis_data,}
                          \hlkwc{iterations} \hlstd{=} \hlnum{1000}\hlstd{,}
                          \hlkwc{family} \hlstd{=} \hlstr{'poisson'}\hlstd{,}
                          \hlkwc{link} \hlstd{=} \hlstr{'log'}\hlstd{,}
                          \hlkwc{priors} \hlstd{=} \hlkwd{list}\hlstd{(}\hlkwc{priormean_intercept} \hlstd{=} \hlnum{0}\hlstd{,}
                                        \hlkwc{priorsd_intercept} \hlstd{=} \hlnum{2}\hlstd{,}
                                        \hlkwc{priormean_slope} \hlstd{=} \hlnum{0.0}\hlstd{,}
                                        \hlkwc{priorsd_slope} \hlstd{=} \hlnum{0.4}\hlstd{,}
                                        \hlkwc{prior_rho_min} \hlstd{=} \hlnum{3}\hlstd{,}
                                        \hlkwc{prior_rho_prob} \hlstd{=} \hlnum{0.01}\hlstd{,}
                                        \hlkwc{prior_sigma_max} \hlstd{=} \hlnum{1}\hlstd{,}
                                        \hlkwc{prior_sigma_prob} \hlstd{=} \hlnum{0.01}\hlstd{))}
\end{alltt}
\end{kframe}
\end{knitrout}

We can get a summary and plot of the model output. The \code{summary} function gives the estimate and standard error of the fixed effect parameters in the model, as well as the negative log likelihood and in-sample performance metrics (root mean squared error, mean absolute error, pearson correlation coefficient and spearman rank correlation coefficient). The \code{plot} function produces two plots: one of the fixed effects parameters and one of the observed data against in-sample predictions, as shown in Figure~\ref{fig:fitmodel}.

\begin{knitrout}
\definecolor{shadecolor}{rgb}{0.969, 0.969, 0.969}\color{fgcolor}\begin{kframe}
\begin{alltt}
\hlkwd{summary}\hlstd{(fitted_model)}
\end{alltt}
\begin{verbatim}
#> Model parameters:
#>                     Estimate Std. Error
#> intercept         -3.1184394  0.2676452
#> slope             -0.3726411  0.1963679
#> slope              0.3031676  0.2073146
#> slope              0.2372611  0.2696821
#> slope             -0.2412063  0.2174388
#> iideffect_log_tau  1.0981493  0.2718976
#> log_sigma         -3.5489432  0.5925067
#> log_rho            0.5596955  0.2984963
#> 
#> Negative log likelihood:  988.313078381041 
#> 
#> In sample performance:
#>       RMSE       MAE pearson spearman log_pearson
#> 1 1.289955 0.9750677       1        1   0.9999994
\end{verbatim}
\end{kframe}
\end{knitrout}

\begin{figure}[!h]
\centering
\begin{knitrout}
\definecolor{shadecolor}{rgb}{0.969, 0.969, 0.969}\color{fgcolor}\begin{kframe}
\begin{alltt}
\hlkwd{plot}\hlstd{(fitted_model)}
\end{alltt}
\end{kframe}
\includegraphics[width=\maxwidth]{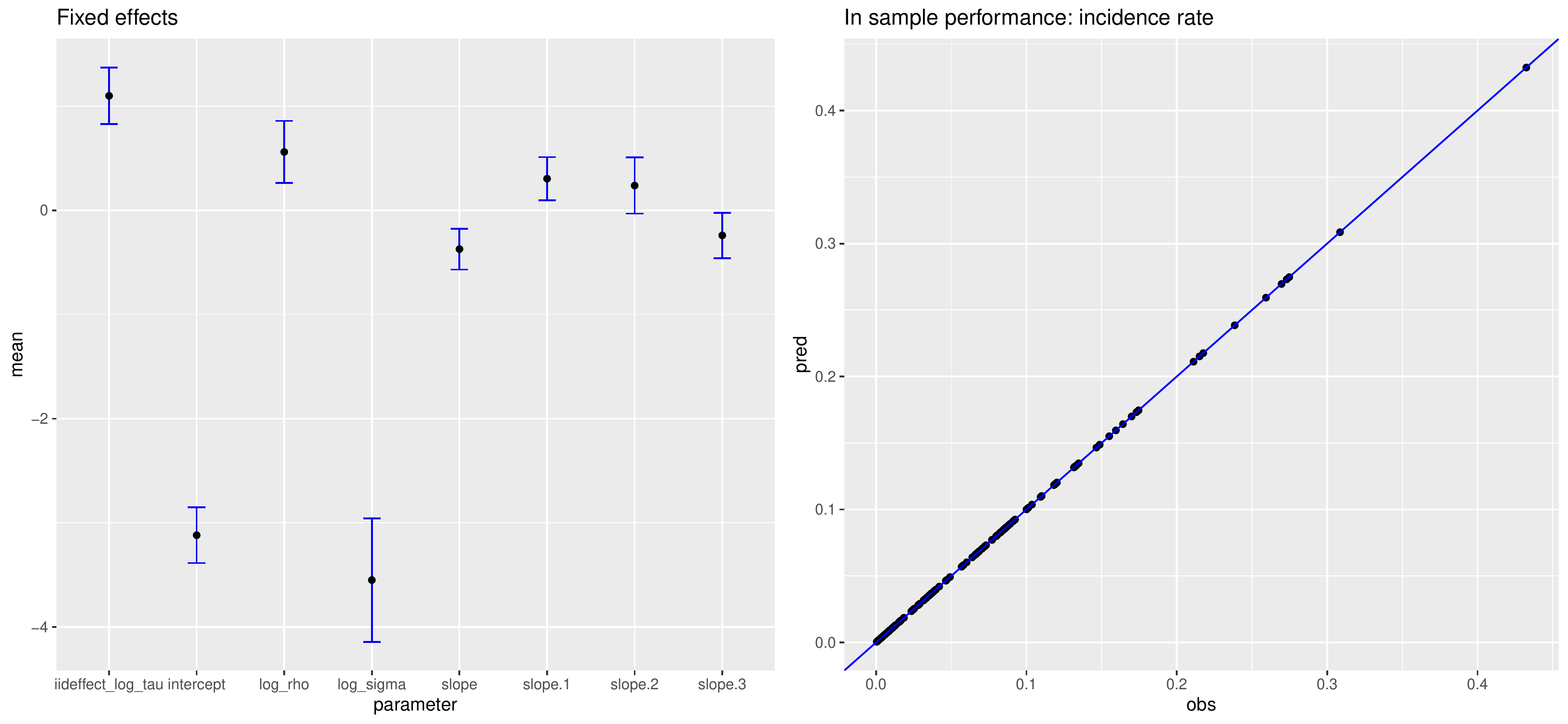} 

\end{knitrout}
\caption{\label{fig:fitmodel} Plot summarising the results of the fitted model. These plots are produced when calling the \code{plot} function on the \code{fit.result} object. The fixed effects plot (left) shows the fitted parameter values with uncertainty estimation for all the fixed effects in the model. The in-sample performance plot (right) shows the predicted incidence rate values for each polygon in the data against the observed values for that polygon in the data.}
\end{figure}

The \code{fit_model} function returns an object of class \code{fit.result}, which is designed to be used directly in the \code{predict} function.
Therefore, now that we have fitted the model, we are ready to predict the malaria incidence rate across Madagascar.

To predict over a different spatial extent to that used in the model, a RasterStack covering the region to make predictions over can be passed as the \code{newdata} argument. If this argument is not given, predictions are made over the covariate rasters used in the fit. If the user wants to include the iid effect from the model as a component in the prediction then the \code{predict_iid} logical flag should be set to TRUE, otherwise, the iid effect will not be predicted. 

For the uncertainty calculations, parameter values are sampled from the posterior distribution and summarised. The number of parameter draws used to calculate the uncertainty is set by the user via the \code{N} parameter (default: 100), and the size of the credible interval (e.g. 75\%, 95\%) to be calculated when summarising is set via the argument \code{CI} (default: 0.95). 

\begin{knitrout}
\definecolor{shadecolor}{rgb}{0.969, 0.969, 0.969}\color{fgcolor}\begin{kframe}
\begin{alltt}
\hlstd{model_predictions} \hlkwb{<-} \hlkwd{predict}\hlstd{(fitted_model)}
\end{alltt}
\end{kframe}
\end{knitrout}

The function \code{predict} returns a list of two objects: the mean predictions (of class \code{predictions}) and the uncertainty rasters (of class \code{uncertainty}). The mean predictions contain a raster of the mean prediction of the incidence rate, as well as rasters of the field, iid (if predicted) and covariate component of the linear predictor. The \code{uncertainty} object contains a RasterStack of the prediction realisations and a RasterStack of the upper and lower credible intervals. 

The \code{plot} function can be used on both the \code{predictions} and \code{uncertainty} objects, as shown in Figures~\ref{fig:subpredictmean} and \ref{fig:predictuncertainty} respectively.

\begin{knitrout}
\definecolor{shadecolor}{rgb}{0.969, 0.969, 0.969}\color{fgcolor}\begin{kframe}
\begin{alltt}
\hlkwd{plot}\hlstd{(model_predictions}\hlopt{$}\hlstd{mean_predictions)}
\end{alltt}
\end{kframe}
\end{knitrout}
\begin{knitrout}
\definecolor{shadecolor}{rgb}{0.969, 0.969, 0.969}\color{fgcolor}\begin{figure}[H]
\subfloat[\label{fig:subpredictmean1}]{\includegraphics[width=0.33\linewidth]{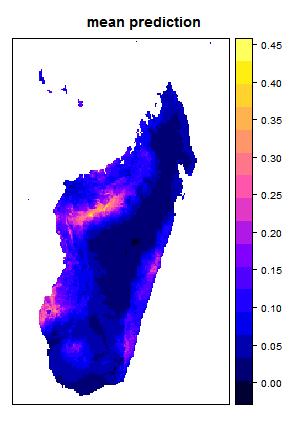} }
\subfloat[\label{fig:subpredictmean2}]{\includegraphics[width=0.33\linewidth]{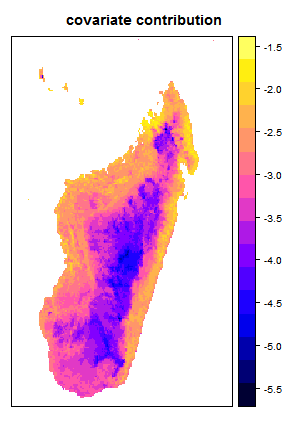} }
\subfloat[\label{fig:subpredictmean3}]{\includegraphics[width=0.33\linewidth]{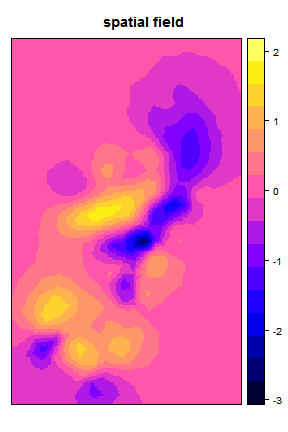} }\caption[Maps of Madagascar showing the fine-scale predictions from the disaggregation model]{Maps of Madagascar showing the fine-scale predictions from the disaggregation model. These maps are produced when calling the plot function on the predictions object. The maps show the (a) mean prediction of malaria incidence rate, (b) covariate contribution to the linear predictor, and (c) field contribution to the linear predictor.}\label{fig:subpredictmean}
\end{figure}

\end{knitrout}

\begin{figure}
\centering
\begin{knitrout}
\definecolor{shadecolor}{rgb}{0.969, 0.969, 0.969}\color{fgcolor}\begin{kframe}
\begin{alltt}
\hlkwd{plot}\hlstd{(model_predictions}\hlopt{$}\hlstd{uncertainty_predictions)}
\end{alltt}
\end{kframe}
\includegraphics[width=\maxwidth]{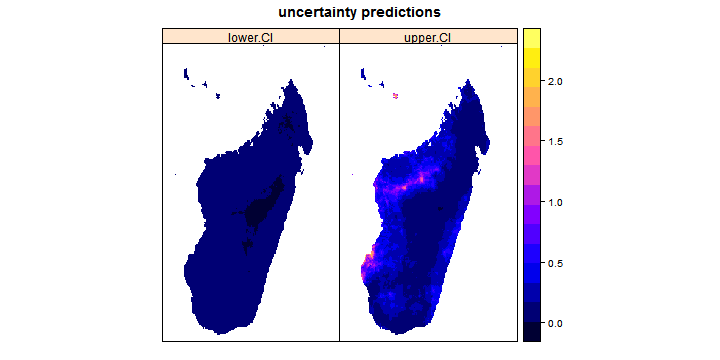} 

\end{knitrout}
\caption{\label{fig:predictuncertainty} Maps of Madagascar showing the fine-scale predictions of lower (2.5\%) and upper (97.5\%) credible intervals of the malaria incidence rate from the disaggregation model. These maps are produced when calling the \code{plot} function on the \code{uncertainty} object.}
\end{figure}

Using three simple functions within the \pkg{disaggregation} package we have been able to fit a Bayesian spatial disaggregation model and predict pixel-level incidence rate across Madagascar using aggregated incidence data and pixel-level environmental covariates.

\newpage
\section{Comparison with Markov chain Monte Carlo (MCMC)} \label{sec:comparison}

In this section we show performance comparisons between modelling using the Laplace approximation provided by the \pkg{disaggregation} package, based on \pkg{TMB}, and using MCMC. Given a function to evaluate the probability density of a distribution at any given point in parameter space, MCMC algorithms construct markov chains to generate samples from this distribution. These algorithms are often slow, particularly in high-dimensional settings, as it can take a long time to converge to and effectively sample from the stationary distribution. The density is evaluated at each step, so this problem is compounded when evaluating this density is computationally expensive. In contrast, the \pkg{disaggregation} package approximates using a Laplace approximation to the posterior to generate posterior samples. This only requires the posterior to be maximised to find the posterior mode and therefore involves relatively few evaluations of the posterior density compared to MCMC techniques, although potentially at the expense of less accurate posterior samples. Here we compare the time and performance of the two techniques.

The model described in Section~\ref{sec:code} for malaria in Madagascar has been optimised using the \pkg{disaggregation} package. Here we fit the same model by running MCMC using the \pkg{tmbstan} package, with the NUTS (no-u-turn sampler) algorithm \citep{hoffman2014no} using four chains. It is important to note that this is a useful feature of the \pkg{disaggregation} package, to be able to create the \pkg{TMB} model object (using the \code{fit_model} function with one iteration) and pass it directly to \pkg{tmbstan}. The model is fitted by running the MCMC algorithm for 8000 iterations with 2000 of those as warmup, which took 94 hours. This number of iterations was chosen by running the MCMC algorithm repeatedly, starting at 1000 iterations, doubling the number of iterations each time until the value of the MCMC convergence statistic, $\hat{R}$, dropped below 1.05 for all model parameters. In contrast, fitting the model using the Laplace approximation via \pkg{TMB} within the \pkg{disaggregation} package took 56 seconds. Fitted parameter values for both of these methods are given in Table~\ref{tab:mcmcresults}. 

\begin{knitrout}
\definecolor{shadecolor}{rgb}{0.969, 0.969, 0.969}\color{fgcolor}\begin{kframe}
\begin{alltt}
\hlkwd{library}\hlstd{(tmbstan)}

\hlstd{tmb_model} \hlkwb{<-} \hlkwd{fit_model}\hlstd{(}\hlkwc{data} \hlstd{= dis_data,}
                       \hlkwc{iterations} \hlstd{=} \hlnum{1}\hlstd{,}
                       \hlkwc{family} \hlstd{=} \hlstr{'poisson'}\hlstd{,}
                       \hlkwc{link} \hlstd{=} \hlstr{'log'}\hlstd{,}
                       \hlkwc{priors} \hlstd{=} \hlkwd{list}\hlstd{(}\hlkwc{priormean_intercept} \hlstd{=} \hlnum{0}\hlstd{,}
                                     \hlkwc{priorsd_intercept} \hlstd{=} \hlnum{2}\hlstd{,}
                                     \hlkwc{priormean_slope} \hlstd{=} \hlnum{0.0}\hlstd{,}
                                     \hlkwc{priorsd_slope} \hlstd{=} \hlnum{0.4}\hlstd{,}
                                     \hlkwc{prior_rho_min} \hlstd{=} \hlnum{3}\hlstd{,}
                                     \hlkwc{prior_rho_prob} \hlstd{=} \hlnum{0.01}\hlstd{,}
                                     \hlkwc{prior_sigma_max} \hlstd{=} \hlnum{1}\hlstd{,}
                                     \hlkwc{prior_sigma_prob} \hlstd{=} \hlnum{0.01}\hlstd{))}

\hlcom{# Running the MCMC algorithm for 94 hours}
\hlstd{start} \hlkwb{<-} \hlkwd{Sys.time}\hlstd{()}
\hlstd{mcmc_out} \hlkwb{<-} \hlkwd{tmbstan}\hlstd{(tmb_model}\hlopt{$}\hlstd{obj,} \hlkwc{chains} \hlstd{=} \hlnum{4}\hlstd{,} \hlkwc{iter} \hlstd{=} \hlnum{8000}\hlstd{,} \hlkwc{warmup} \hlstd{=} \hlnum{2000}\hlstd{,}
                    \hlkwc{cores} \hlstd{=} \hlkwd{getOption}\hlstd{(}\hlstr{'mc.cores'}\hlstd{,} \hlnum{4}\hlstd{))}
\hlstd{end} \hlkwb{<-} \hlkwd{Sys.time}\hlstd{()}
\hlkwd{print}\hlstd{(end} \hlopt{-} \hlstd{start)}

\hlcom{# Plot the trace of the parameter values for each sampling method}
\hlkwd{stan_trace}\hlstd{(mcmc_out,}
           \hlkwc{pars} \hlstd{=} \hlkwd{c}\hlstd{(}\hlstr{'intercept'}\hlstd{,}
                    \hlstr{'slope[1]'}\hlstd{,} \hlstr{'slope[2]'}\hlstd{,} \hlstr{'slope[3]'}\hlstd{,} \hlstr{'slope[4]'}\hlstd{,}
                    \hlstr{'iideffect_log_tau'}\hlstd{,} \hlstr{'log_sigma'}\hlstd{,} \hlstr{'log_rho'}\hlstd{))}
\end{alltt}
\end{kframe}
\end{knitrout}
\begin{table}
\centering
\begin{tabular}{l|ll|ll}
 & \multicolumn{2}{|c|}{MCMC (94 hours)} & \multicolumn{2}{|c}{TMB (56 seconds)} \\
Parameter & Mean & SD & Mean & SD \\
\hline
Intercept & -3.09 & 0.33 & -3.12 & 0.27 \\
Slope 1 & 0.20 & 0.28 & 0.24 & 0.27 \\
Slope 2 & 0.32 & 0.21 & 0.30 & 0.21\\
Slope 3 & -0.36 & 0.20 & -0.37 & 0.20 \\
Slope 4 & -0.24 & 0.23 & -0.24 & 0.21 \\
$log(\tau_u)$ & 1.08 & 0.27 & 1.10 & 0.27 \\
$log(\sigma)$ & -3.34 & 0.59 & -3.55 & 0.59 \\
$log(\rho)$ & 0.72 & 0.31 & 0.56 & 0.30 \\
\end{tabular}
\caption{Fitted model parameter values using both MCMC and using \pkg{TMB} within the \pkg{disaggregation} package.}
\label{tab:mcmcresults}
\end{table}

\begin{figure}[t!]
\centering
\includegraphics[width=\linewidth]{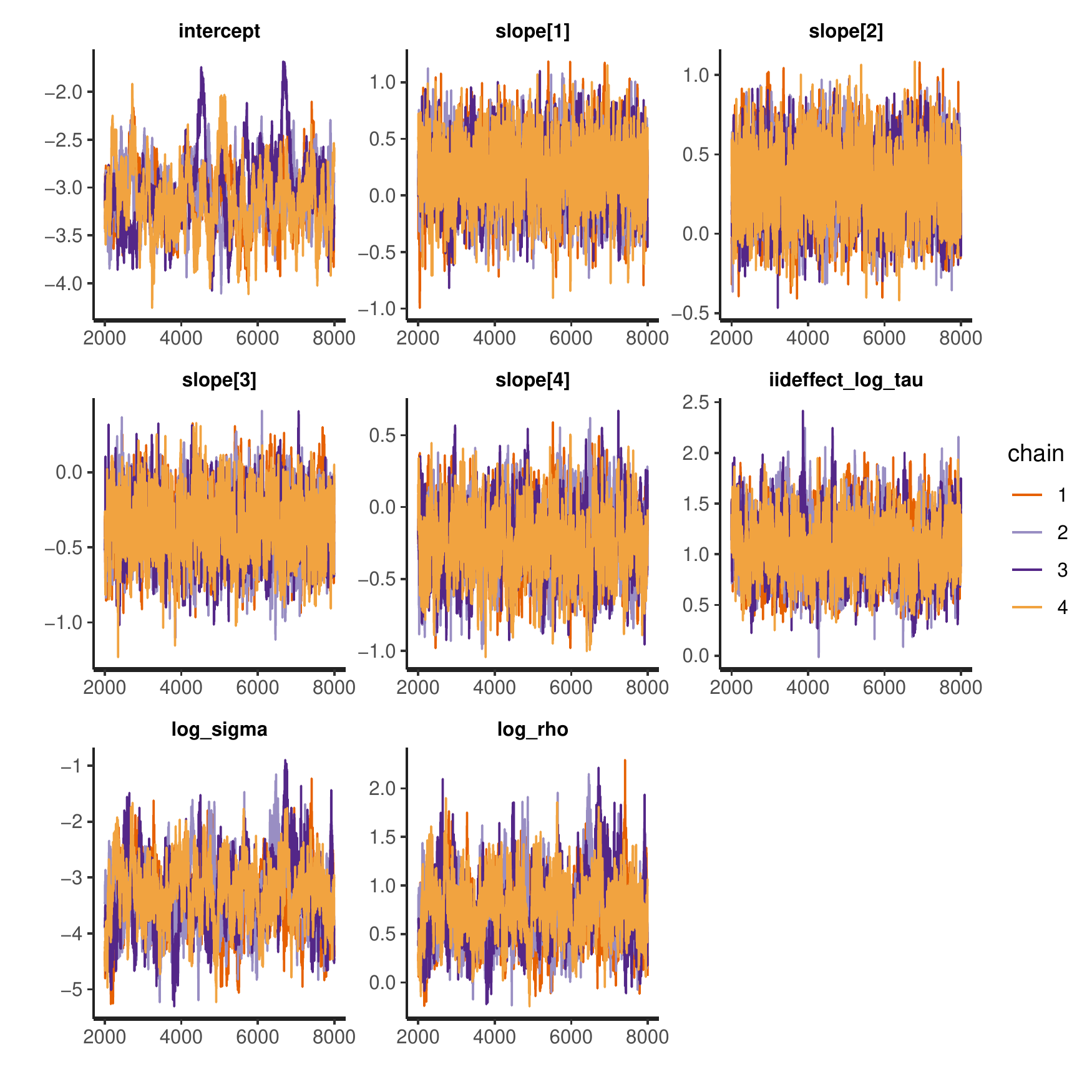}
\caption{Trace of the fixed effects parameters for MCMC using NUTS sampling, running the algorithm for 94 hours. It was run for 8000 iterations with 2000 of those as warmup.}
\label{fig:mcmclong}
\end{figure}

The trace of the MCMC parameter values is given in Figure~\ref{fig:mcmclong}. It can be seen that the MCMC algorithm has been run for long enough to get sufficient chain mixing. The \pkg{disaggregation} package produces similar results to the MCMC algorithm. However the model fitting using the \pkg{disaggregation} package took 56 seconds in contrast to the 94 hours taken for the MCMC run. Therefore, it can be seen that the \pkg{disaggregation} package provides a quick and simple way to run disaggregation models, that can be prohibitively slow using MCMC.

%


\section{Conclusions} \label{sec:conclusions}

Disaggregation modelling, which involves fitting models at fine-scale resolution using areal data over heterogenous regions, has become widely used in fields such as epidemiology and ecology. The \pkg{disaggregation} package implements Bayesian spatial disaggregation modelling with a simple, easy to use \proglang{R} interface. The package includes simple data preparation, fitting and prediction functions that allow some user-defined model flexibility. In this paper we have presented an application of the package, predicting malaria incidence rate across Madagascar from aggregated count data and environmental covariates. 

The modelling framework is implemented using the Laplace approximation and automatic differentiation within the \pkg{TMB} package. This allows fast, optimised calculations in C++. These disaggregation models are computationally intensive and take a long time using MCMC optimisation techniques. Using \pkg{TMB}, the models are much faster and produce similar results.

Future work could be done extending the \pkg{disaggregation} package to include spatio-temporal disaggregation models. This would require a spatio-temporal field as well as dynamic covariates, and would be significantly more computationally intensive. Additionally, tools for cross-validation could be included within the package. Cross-validation of spatial models is non-trivial due to the spatial autocorrelation in the data. 

The \pkg{disaggregation} package provides a simple, useful interface to perform spatial disaggregation modelling, with reasonable flexibility, as well as having the scope to be extended to more complex disaggregation models.


\section*{Computational details}

The results in this paper were obtained using
\proglang{R}~3.6.1 with the
\pkg{TMB}~1.7.15 package. \proglang{R} itself
and all packages used are available from the Comprehensive
\proglang{R} Archive Network (CRAN) at
\url{https://CRAN.R-project.org/}, apart from \pkg{INLA}, which can be 
installed in \proglang{R} using the command:
\begin{knitrout}
\definecolor{shadecolor}{rgb}{0.969, 0.969, 0.969}\color{fgcolor}\begin{kframe}
\begin{alltt}
\hlkwd{install.packages}\hlstd{(}\hlstr{"INLA"}\hlstd{,}
                 \hlkwc{repos} \hlstd{=} \hlkwd{c}\hlstd{(}\hlkwd{getOption}\hlstd{(}\hlstr{"repos"}\hlstd{),}
                           \hlkwc{INLA}\hlstd{=}\hlstr{"https://inla.r-inla-download.org/R/stable"}\hlstd{),}
                 \hlkwc{dep}\hlstd{=}\hlnum{TRUE}\hlstd{)}
\end{alltt}
\end{kframe}
\end{knitrout}

\section*{Acknowledgments}

We would like to thank the Bill and Melinda Gates Foundation for funding this research.


\bibliography{refs}

\end{document}